# Information-Theoretic Approach to Strategic Communication as a Hierarchical Game

Emrah Akyol, Cédric Langbort, Tamer Başar
Coordinated Science Laboratory, University of Illinois at Urbana-Champaign
Email: {akyol, langbort, basar1}@illinois.edu

*Abstract*—This paper analyzes the information disclosure problems originated in economics through the lens of information theory. Such problems are radically different from the conventional communication paradigms in information theory since they involve different objectives for the encoder and the decoder, which are aware of this mismatch and act accordingly. This leads, in our setting, to a hierarchical communication game, where the transmitter announces an encoding strategy with full commitment, and its distortion measure depends on a private information sequence whose realization is available at the transmitter. The receiver decides on its decoding strategy that minimizes its own distortion based on the announced encoding map and the statistics. Three problem settings are considered, focusing on the quadratic distortion measures, and jointly Gaussian source and private information: compression, communication, and the simple equilibrium conditions without any compression or communication. The equilibrium strategies and associated costs are characterized. The analysis is then extended to the receiver side information setting and the major changes in structure of optimal strategies are identified. Finally, an extension of the results to the broader context of decentralized stochastic control is presented.

## I. Introduction

The field of information economics (IE) has considered several different models of "communication" (information transmission) between two agents, a transmitter and a receiver, with different objectives, see for example the well-known strategic information transmission (SIT) and information disclosure models [3], [4], [6].

The conclusions of such studies (especially the structure of transmitter/receiver policies achieving equilibrium) are particularly interesting from an information-theoretic (IT) viewpoint, given the relative dearth of results in this field related to mismatched goals between the transmitter and the receiver. Indeed, most IT models to date that have considered source compression with mismatched distortion measures have done so in a context where mismatch was caused either by Nature (as in the worst-case or robust design approaches of [7], [8]) or by a secondary adversarial decoder [9]. On the other hand, communication models in IE do not consider compression or actual physical channel which are of main interest in IT [10]. This work studies the interplay of the game and the communication aspects of such strategic communication models in IE and IT.

More specifically, this work focuses on compression and communication scenarios where the better informed transmitter communicates with a receiver who makes the ultimate decision concerning both agents. The objectives of the transmitter and the receiver are only partially aligned. As opposed to the classical work in economics, which is strategic information transmission [6], we consider the problem as a Stackelberg game [11] where the transmitter is committed to its policy[1] ex-ante, and the receiver (the follower) is aware of the transmitter (the leader) policy and optimizes it as a function of the encoding policy and the statistics. In this sense, our model is along the lines of some recent works in IE, namely information disclosure [4] and Bayesian persuasion [3]. This framework is similar to the conventional communication settings analyzed in IT (see e.g., [10]), in terms of the commitment based on statistics (not realization) as opposed to [6]. Hence, it enables the use of Shannon theoretic arguments [10] to derive fundamental limits of compression and communication.

As a side note, we note that very recently the optimality of linear policies for a generalized version of this problem was independently shown in the working paper [5], using tools from convex optimization. We also note that Stackelberg games with quadratic objectives and Gaussian variables (which will be referred to as quadratic-Gaussian (Q-G) setting throughout the paper) have been well studied in the control literature, see e.g., [12]. However, the problems we consider here involve communication (and hence, in a sense, non-classical information structures [13]) and are therefore fundamentally different from the ones in control without any communication. Control problems with communication (cf. [14]), particularly ones with Gaussian channels have been analyzed in non-strategic settings employing the Shannon bounds. However, there are no known Shannon-like bounds in the strategic settings, and this is indeed the main subject of this paper.

Highlights of the contributions of this paper are as follows:

- We show, using statistical tools widely used in information theory, that linear policies are *uniquely optimal* for the Q-G simple equilibrium (without any compression or noise in the communication).

Parts of the material in this paper were presented at the IEEE Information Theory Workshop, Korea, Oct. 2015 [1] and the IEEE International Symposium on Information Theory, Spain, July 2016 [2]. This work was supported in part by AFOSR MURI Grant FA9550-10-1-0573. Authors would like thank to Joel Sobel for valuable comments and encouragements, and informing us on the relationship between our model and the one in [3]. We also thank an anonymous reviewer for pointing out the relevance of [3], [4] and particularly [5], and the editor for valuable comments that helped us clarify and streamline the presentation of our results.

---
[1]In this paper, we use "policy" and "strategy" interchangeably.

- We determine the single-letter characterization of the fundamental limits of strategic compression, and explicitly compute this quantity for the Q-G setting.
- We show optimality of single-letter linear strategies in the Q-G communication setting. This result parallels the well-known optimality of single-letter mappings for the same setting, but without the strategic aspect of the problem [15].
- We analyze the impact of the receiver side information on the structure of the results.
- Finally, we demonstrate the use of information theoretic results in strategic decision making/control problems involving the Gaussian test channel.

## II. Preliminaries

### A. Notation

$\mathbb{R}$ and $\mathbb{R}^+$ denote the respective sets of real numbers and positive real numbers. We let $\mathbb{E}(\cdot)$ denote the expectation operator. The Gaussian density with mean $\mu$ and variance $\sigma^2$ is denoted by $\mathcal{N}(\mu, \sigma^2)$. All logarithms in the paper are natural logarithms, and the integrals are, in general, Lebesgue integrals. $\mathcal{S}$ denotes the set of Borel measurable, square integrable functions $\{f : \mathbb{R} \to \mathbb{R}\}$. All the alphabets used in this paper are the real line, but for clarity we denote them by separate letters $\mathcal{X}$, $\theta$, and $\hat{\mathcal{X}}$ for the source, the private information, and the reconstruction, respectively. We use standard information theoretic and game theoretic notations for the related results throughout this paper (cf. [11], [16]).

### B. Overview of Communication Games in Economics

There exists a vast amount of literature in economics on communicating information in transmitter-receiver games: using as advertising [17], [18], education [19], disclosure of verifiable information [20], or cheap talk [6], or information disclosure [3], [4]. Here, we review two distinct, well-known models: The cheap talk model of [6] and the more relevant information disclosure model in [4] and [3].

*1) Strategic Information Transmission (Cheap Talk):* In the SIT model [6], there are two players: a transmitter and a receiver. The state information, $X$, is drawn from a population with density $f(\cdot)$ with bounded support, say $[0,1]$. $X$ is available only to the transmitter. The receiver, based on the transmitter output, takes action $a \in \mathbb{R}$. The utility functions of the transmitter and the receiver are respectively $U_T(x,a,b)$ and $U_R(x,a)$, where $b$ is a deterministic *bias parameter* that measures the differences in the preferences of the agents. All aspects of the game, except the realization of $X$, are common knowledge.

The game is described as follows: the transmitter observes the source output $X$ and transmits a message $y \in \mathcal{Y}$, where $\mathcal{Y}$ is an infinite set. The receiver observes $y$ and chooses an action $a \in \mathbb{R}$ which determines the pay-offs. A pure strategy[2] equilibrium comprises an encoding strategy $g^* : [0,1] \to \mathcal{Y}$ and a decoding (action) strategy $h^* : \mathcal{Y} \to \mathbb{R}$ (hence, $a = h(Y)$) such that

$$g^*(X) = \underset{g}{\arg\max}\, U_T(X, h^*(g(X)), b)$$
$$h^*(Y) = \underset{h}{\arg\max}\, \mathbb{E}\{U_R(X, h(Y))|Y\} \quad (1)$$

where $Y = g^*(X)$. Note that the pair $(g^*, h^*)$ constitutes a Nash equilibrium for the underlying two-player game. We also note that such a pair is not necessarily unique. In fact, there may exist multiple Nash equilibria, each one leading to a different cost pair; see e.g., [21] on the selection of the "best" equilibrium among these. An interesting aspect is that there is no cost associated with the message $y$, which is why this setting has also been referred to as "cheap talk" [22]. The main result of [6] states that the mappings at the equilibrium have structure:

**Theorem 1** ([6]). *Any $g^*(X)$ and $h^*(Y)$ that satisfy (1) are non-injective (quantizer based) mappings.*

This result has several interesting aspects. First, in contrast with the classical communication setting, where quantization is imposed by rate constraint or channel noise, here quantization occurs solely because of the mismatch between agents' objectives. Second, this structural result holds for any arbitrary source[3], even for the Q-G setting. Third, a frequently exploited result in estimation and control theory states optimality of linear strategies for the Q-G setting, see e.g., [24]; hence it is surprising to see that such an optimal structure breaks down in a strategic setting.

An important aspect of the SIT model is that the transmitter is not committed to any mappings before seeing the realization of $X$, i.e., the transmitter decides on the mapping $g(\cdot)$ after seeing the realization of $X$. Hence, a Nash equilibrium is sought among all transmitter/receiver strategies. While the conclusions of the SIT model are very interesting for an information theorist (e.g., the main mathematical tool in lossy source coding, *the quantizer*, arises entirely from strategic aspect of the problem), the model does not apply to settings in communication where the transmitter is committed to its encoding map which is designed purely based on statistics and used repeatedly.

*2) Overview of Information Disclosure Model (Bayesian Persuasion):* Consider now the Stackelberg solution as opposed to the Nash equilibrium considered in [6] (keeping all other features of the cheap talk game described above unaltered), where the transmitter is the leader and the receiver is the follower. The game proceeds as follows: the transmitter plays first and announces an encoding mapping. As opposed to the game in [6], the transmitter is *committed* to its encoding mapping, i.e., the transmitter cannot change it after the receiver plays. The receiver, knowing this commitment, determines its own mapping that maximizes its pay-off, given the encoding mapping. The transmitter, of course, will anticipate this, and pick its mapping accordingly.

---
[2] Limitation to pure strategies does not introduce any loss of generality here.

[3] When the distortion measures are quadratic, the technical requirement on the source statistics, that is having a density with bounded support, can be removed; see e.g., [23]

In [3], this setting is studied with scalar state and mappings, and conditions for which full or no disclosure are optimal are characterized. In [4], the same problem setting is studied with two-dimensional state and private information available only to the receiver. Both of these works analyze the optimal strategies for discrete variables. In [5], [25], noiseless Stackelberg equilibrium involving Gaussian variables and quadratic payoff functions is studied.

### C. Setting-1: Simple (Noiseless) Equilibrium

Here we introduce one of the problems studied in this paper. Consider the general communication system whose block diagram is depicted in Figure 1. The source $X$ and private information $\theta$ are mapped into $U \in \mathbb{R}$ which is fully determined by the conditional distribution $p(\cdot|x,\theta)$. For the sake of brevity, and with a slight abuse of notation, we refer to this as a stochastic mapping $U = g(X, \theta)$, so that

$$\mathbb{P}(g(X,\theta) \in \mathcal{U}) = \int_{u' \in \mathcal{U}} p(u'|x,\theta) \mathrm{d}u' \quad \forall \mathcal{U} \subseteq \mathbb{R} \quad (2)$$

holds almost everywhere in $X$ and $\theta$. Let the set of all such mappings be denoted by $\Gamma$ (which has a one-to-one correspondence to the set of all the conditional distributions that construct the transmitter output $U$).

In the most general communication setting (see Sections II-E and III-C), we consider an additive noise channel as shown in Figure 1, with Gaussian noise $N \sim \mathcal{N}(0, \sigma_N^2)$, hence the input to the receiver is $Y = U + N$. Here, however we focus on the simpler problem where there is no channel noise, i.e., we effectively assume $\sigma_N^2 = 0$, and hence $Y = U$ (almost everywhere). The receiver produces an estimate of the source $\hat{X}$ through a mapping $h \in \mathcal{S}$ as $\hat{X} = h(Y)$. The objective of the receiver is to pick $h \in \mathcal{S}$ so as to minimize

$$D_D = \mathbb{E}\{d_D(X, \hat{X})\} \quad (3)$$

where $d_D : \mathcal{X} \times \hat{\mathcal{X}} \to \mathbb{R}^+$ is the distortion measure associated with the receiver's distortion. The objective of the transmitter is to minimize

$$D_E = \mathbb{E}\{d_E(X, \theta, \hat{X})\} \quad (4)$$

using the freedom in the choice of the mapping $g(\cdot,\cdot) \in \Gamma$, given that the receiver chooses $h \in \mathcal{S}$ as above, where $d_E : \mathcal{X} \times \theta \times \hat{\mathcal{X}} \to \mathbb{R}^+$ is the transmitter's distortion metric. In the following, we present this optimization problem formally:

**Problem 1.** *Find essentially unique*[4] $g^*(\cdot, \cdot) \in \Gamma$ *and* $h^*(\cdot) \in \mathcal{S}$ *that satisfy*

$$g^*(X, \theta) = \underset{g \in \Gamma}{\arg\min}\, d_E(X, \theta, h^*(g(X,\theta)))$$
$$h^*(Y) = \underset{h \in \mathcal{S}}{\arg\min}\, \mathbb{E}\{d_D(X, h(Y)|Y\}$$

[4]In this noiseless setting, the encoding/decoding mappings $F(g(X,\theta))$ and $F^{-1}(h(Y))$ yield the same costs as $g(X,\theta)$ and $h(Y)$, where $F(\cdot)$ is any invertible function. This function corresponds to different permutations of labels if the message space is finite as assumed in most prior work in IE [3], [4], [6]. To account for such trivially equivalent pairs of mappings, we use the term "essentially unique".

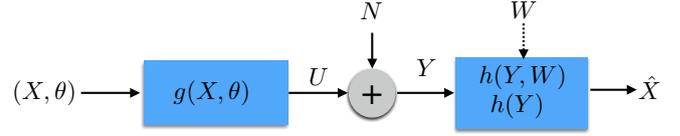

Fig. 1: The strategic variant of the Gaussian test channel, with or without receiver side information $W$.

where $Y = g(X, \theta)$.

**Quadratic-Gaussian Setting**: Most of our results concern the setting where the source and the private information are jointly Gaussian i.e., $(X, \theta) \sim \mathcal{N}(0, R_{X\theta})$ where, without any loss of generality, $R_{X\theta}$ is parametrized as $R_{X\theta} = \sigma_X^2 \begin{bmatrix} 1 & \rho \\ \rho & r \end{bmatrix}$, with $r > \rho^2$, and the distortion measures are given as follows:

$$d_E(x,\theta,y) = (x + \theta - y)^2; \quad d_D(x,y) = (x-y)^2. \quad (5)$$

Hence, in the Q-G setting, we have the following cost functions:

$$D_E = \mathbb{E}\left\{(X + \theta - \hat{X})^2\right\}; \quad D_D = \mathbb{E}\left\{(X - \hat{X})^2\right\}. \quad (6)$$

**Remark 1.** *The Q-G setting formulation (without any noise or compression) can be viewed as a special case of the problem analyzed in [5]. However, as shown later in Theorem 4, the information-theoretic approach taken in this paper provides more conclusive results, namely the uniqueness of the optimal strategies, for this noiseless setting, and also allows for an extension of the analysis to noisy settings.*

Our result regarding this equilibrium is that, in sharp contrast with the original SIT which considers the Nash equilibrium, the Stackelberg solution for the Q-G setting entails linear encoding-decoding strategies as the *essentially unique optimal strategy pair* (see Theorem 4). While the existence of a linear optimal strategy can be shown using tools from convex optimization (see [5]), proving the essential uniqueness of the optimal policy poses a significant challenge, which can be circumvented using two tools from probability theory. In the following, we present an overview of these powerful tools.

*1) Functional Representation Lemma:* Like many ideas in information theory, the functional representation lemma has its roots in Shannon's original pioneering work [26]. Since then, different variations of this lemma have been used for different problems in information theory, see e.g., [27]. The following variation can be found in [28, page 626].

**Lemma 1.** *For a given set of random variables $X, Y, W$, there exist a random variable $Z$ distributed independent of $Y$, and a deterministic function $\gamma$ such that $W$ can be expressed as*

$$W = \gamma(Y, Z)$$

*and $X - (Y, W) - Z$ forms a Markov chain in this order.*

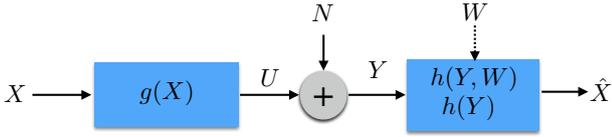

Fig. 2: The (non-strategic) Gaussian test channel, with or without receiver side information $W$.

*2) Maximal Correlation Coefficient:* The maximal correlation coefficient between two random variables[5] $X$ and $Y$, introduced in [29], is

$$f_m(X,Y) = \sup \mathbb{E}\{f(X)g(Y)\} \quad (7)$$

where the supremum is taken over all (Borel) measurable functions $f, g$ with

$$\mathbb{E}\{f(X)\} = \mathbb{E}\{g(Y)\} = 0, \quad (8)$$
$$\mathbb{E}\{f^2(X)\} = \mathbb{E}\{g^2(Y)\} = 1. \quad (9)$$

The following is a well-known result, see e.g., [30] for a proof.

**Lemma 2.** *For jointly Gaussian random variables $\xi_1$ and $\xi_2$,*

$$f_m(\xi_1, \xi_2) = |\mathbb{E}\{\xi_1 \xi_2\}| \quad (10)$$

*where the supremum in $f_m$ is achieved uniquely by linear (identity) mappings.*

Due to Lemma 2 and the tensorization[6] property (shown in [31]), maximal correlation has played a significant role in several problems in information theory, see e.g., [32].

*D. Setting-2: Strategic Compression*

Let us first present an overview of the basic (non-strategic) results in source compression. Assume a memoryless source and a single-letter, bounded, and additive distortion measure $d : \mathcal{X} \times \mathcal{Y} \longrightarrow \mathbb{R}^+$, *i.e.*,

$$d(X^n, Y^n) = \frac{1}{n} \sum_{t=1}^{n} d(X_t, Y_t). \quad (11)$$

A block code pair $(f_E, f_D)$ consists of an encoding function $f_E : \mathcal{X}^n \longrightarrow \mathcal{M}$ which maps the source to index set $\mathcal{M}$, and a decoding function $f_D : \mathcal{M} \longrightarrow \mathcal{Y}^n$. A rate-distortion pair $(R, D)$ is called *achievable* if for every $\delta > 0$ and sufficiently large $n$, there exists a block code $(f_E, f_D)$ such that

$$\frac{1}{n} \log |\mathcal{M}| \leq R + \delta$$
$$E\{d(X^n, f_D(f_E(X^n)))\} \leq D + \delta.$$

The fundamental result in information theory [16] is that the rate-distortion (R-D) function expressing the minimum achievable rate $R$ for a prescribed level of distortion $D$, denoted as $R(D)$, is given by minimizing the mutual information $I(X;Y)$ over all conditional distributions $P_{Y|X}(y|x)$ that maintain the prescribed distortion:

$$R(D) = \inf_{P_{Y|X}(y|x): E\{d(X,Y)\} \leq D} I(X;Y). \quad (12)$$

Now, we define the strategic compression problem similar to its non-strategic counterpart where a memoryless source $X^n$ and a private information sequence $\theta^n$ are mapped to an index set $\mathcal{M}$ by $f_E : \mathcal{X}^n \times \theta^n \longrightarrow \mathcal{M}$. The decoder applies $f_D : \mathcal{M} \longrightarrow \mathcal{Y}^n$ to generate the reconstruction sequence $\hat{X}^n$. An achievable triple $(R, D_E, D_D)$ satisfies

$$\frac{1}{n} \log |\mathcal{M}| \leq R + \delta$$
$$E\{d_E^n(X^n, \theta^n, f_D(f_E(X^n, \theta^n)))\} \leq D_E + \delta$$
$$E\{d_D^n(X^n, f_D(f_E(X^n, \theta^n)))\} \leq D_D + \delta,$$

for every $\delta > 0$ and sufficiently large $n$. The set of achievable triples $(R, D_E, D_D)$ is denoted here as $\mathcal{RD}_S$.

**Problem 2.** *Find the equilibrium points $(R, D_E, D_D)$ in $\mathcal{R}_{DS}$ achieved by the mappings that satisfy:*

$$f_E^*(X^n, \theta^n) = \underset{f_E}{\mathrm{argmin}}\, \mathbb{E}\{d_E^n(X^n, \theta^n, f_D^*(f_E(X^n, \theta^n)))\}$$
$$f_D^*(f_E(X^n, \theta^n)) = \underset{f_D}{\mathrm{argmin}}\, \mathbb{E}\{d_D^n(X^n, f_D(f_E(X^n, \theta^n)))\}$$

*E. Setting-3: Gaussian Test Channel and Equilibrium with Noisy Channel*

Consider the general communication system whose block diagram is shown in Figure 2, where the source $X \sim \mathcal{N}(0, \sigma_X^2)$ is to be transmitted to the receiver via $g \in \mathcal{S}$ as $U = g(X)$ over an additive Gaussian channel; hence the input to the receiver is $Y = U + N$, where $N \sim N(0, \sigma_N^2)$ is statistically independent of $X$. The receiver produces its output $\hat{X}$ through an $h \in \mathcal{S}$ as $\hat{X} = h(Y)$[7]. The objective of both agents is to minimize $\mathbb{E}\{(X - \hat{X})^2\}$, while the transmitter has an average power constraint $\mathbb{E}\{U^2\} \leq P_T$.

Although this problem is formulated in a single-letter setting, the solution is obtained by expanding the feasible solution space to $n$-letter strategies, i.e., by solving the information theoretic version of the problem (*Shannon sense optimality*). Obviously allowing more delay, the Shannon sense optimal solution should perform at least as well as the zero-delay one. The following well-known result of Goblick [15] states that these solutions are in fact identical.

**Theorem 2** ([15]). *For the Gaussian test channel problem, single-letter mappings*

$$g(X) = \sqrt{\frac{P_T}{\sigma_X^2}} X, \quad h(Y) = \frac{\sigma_X^2}{P_T + \sigma_N^2} \sqrt{\frac{P_T}{\sigma_X^2}} Y$$

*are the essentially unique[8], Shannon sense optimal encoding/decoding mappings.*

---
[5]We assume, without any loss of generality, that the variables are zero-mean.

[6]A measure of dependence $\Phi(X, Y)$ is said to have tensorization property if for any $n$ i.i.d. tuples of $X$ and $Y$, $\Phi(X^n, Y^n) = \Phi(X, Y)$.

[7]Here, by a slight abuse of notation, we take $g$ and $h$ belong to the same set of mappings, $\mathcal{S}$, which is the set of all Borel measurable, square integrable functions.

[8]If the pair $g(X) = cX$ and $h(Y) = dY$ is a solution to this problem, the pair $g(X) = -cX$ and $h(Y) = -dY$ is also a solution due to symmetry, which is why the solution is "essentially" unique.

**Remark 2.** *This optimality breaks down in the presence of receiver side information (a situation which will be referred to as SI throughout the paper), shown as $W$ in Figure 2, and in that case, linear strategies are no longer optimal even in the set of zero-delay strategies (see e.g., [33]). This observation highlights the fact that conclusions on optimality of linear strategies in noiseless settings do not directly carry over to noisy settings. As we will analyze in Section IV, the strategic aspect brings up cases where an optimality result, similar to the one in Theorem 2 holds in the strategic SI setting, depending on the problem parameters.*

In Section III-C, we investigate whether such a result holds also for the *strategic* variant of the same problem depicted in Figure 1. We refer to this setting as the "noisy equilibrium" (see Section III-C) since this is essentially a noisy version of the problem described in Section II-C. In the following, we formalize this problem.

**Problem 3.** *Find $g^*(\cdot, \cdot) \in \Gamma$ and $h^*(\cdot) \in \mathcal{S}$ that satisfy*

$$g^*(X, \theta) = \underset{g \in \Gamma}{\operatorname{argmin}} \, \mathbb{E}\left\{d_E(X, \theta, h^*(Y))\right\}$$
$$h^*(Y) = \underset{h \in \mathcal{S}}{\operatorname{argmin}} \, \mathbb{E}\left\{d_D(X, h(Y))\right\}$$

*where $Y = g(X, \theta) + N$, $N \sim \mathcal{N}(0, \sigma_N^2)$, and $d_E$ and $d_D$ are as given in (5).*

### F. Decentralized Stochastic Control Problems

The optimality of linear strategies plays a central role in many problems in control and economics, particularly in team decision theory. While the solutions to linear, quadratic and Gaussian (LQG) team problems with classical or quasi-classical information structures are well known to be linear, when the information structure is non-classical, such problems may or may not admit linear (or affine) optimal solutions (cf. [14]). The celebrated 1968 counterexample of Witsenhausen, where optimal solution is not affine, belongs to this family of problems [34]. Another example (of a problem with non-classical information) is the Gaussian test channel, which however admits a linear optimal solution (see Section II-E). In the following, we present a generic setting associated with these problems.

Consider the communication setting depicted in Figure 2 (without SI). All variables are Gaussian and the agents operate through the mappings $g, h \in \mathcal{S}$. The common objective of both agents is minimization of $J = \mathbb{E}\{\varphi(X, U, \hat{X})\}$: an expectation of a function $\varphi$ in the form of a second-order polynomial of $X, U$ and $\hat{X}$, over the mappings $g, h$, where expectation is over all random quantities involved. The Gaussian test channel, which admits a linear optimal solution, corresponds to $\varphi = (X - \hat{X})^2 + k_1 U^2$ with $k_1 > 0$ (see Theorem 2). The counterexample of Witsenhausen, which corresponds to $\varphi = (X + U - \hat{X})^2 + k_1 U^2$ with $k_1 > 0$, still admits an optimal solution but it is not linear [34]. In [13], these two results have been generalized to obtain the conditions on $\varphi$ which guarantee that the problem admits a linear (or affine) optimal solution.

**Theorem 3** ([13]). *The problem admits a linear optimal solution if, and only if, $\varphi$ does not involve any product term between $U$ and $\hat{X}$.*

In Section V, we extend this result, which can be viewed as a generalization of the Gaussian test channel, to strategic settings. More formally, we have the following problem.

**Problem 4.** *Find $g^*(\cdot, \cdot) \in \Gamma$ and $h^*(\cdot) \in \mathcal{S}$ that satisfy*

$$g^*(X, \theta) = \operatorname{argmin} \mathbb{E}\{\varphi_E(X, \theta, U, \hat{X})\} \quad (13)$$
$$h^*(Y) = \operatorname{argmin} \mathbb{E}\{\varphi_D(X, \theta, U, \hat{X})\}$$

*where $Y = g(X, \theta) + N$, $N \sim \mathcal{N}(0, \sigma_N^2)$, and $\varphi_E$ and $\varphi_D$ are second-order polynomials of $X, \theta, U, \hat{X}$.*

## III. MAIN RESULTS

### A. Equilibrium Conditions

We first characterize the Q-G equilibrium with the noiseless channel.

**Theorem 4.** *In the noiseless Q-G setting, the essentially unique solution to Problem 1 is given as $g^*(X, \theta) = X + \alpha \theta$ and $h^*(Y) = \kappa Y$, where $\alpha$ and $\kappa$ are constants given as:*

$$\alpha = \frac{A - 1}{2(r + \rho)}, \quad \kappa = \frac{1 + \alpha \rho}{1 + \alpha^2 r + 2\alpha \rho} \quad (14)$$

*Costs at this Stackelberg solution are*

$$D_E = \sigma_X^2 \left(1 + \frac{(A - 3)(r + \rho)}{A - 1}\right) \quad (15)$$

$$D_D = \sigma_X^2 \left(\frac{(r - \rho^2)(A - 1)}{A(2r + A\rho + \rho)}\right) \quad (16)$$

*where $A = \sqrt{1 + 4(r + \rho)}$.*

*Proof.* The optimal decoding mapping is $h(Y) = \mathbb{E}\{X|Y\}$ regardless of the choice of encoder's policy $g$. Hence, the problem simplifies to an optimization over the encoding mapping $g$. Consider the dual (equivalent) problem of minimizing $D_E$ subject to a fixed $D_D$. Expanding $D_E$, we have

$$D_E = \mathbb{E}\{(X - \mathbb{E}\{X|Y\})^2\} + 2\mathbb{E}\{\theta(X - \mathbb{E}\{X|Y\})\} + \mathbb{E}\{\theta^2\}.$$

Noting that the first term is $D_D$, and the last term is $\mathbb{E}\{\theta^2\}$ (constant with respect to the optimization variables), the problem simplifies to minimizing $\mathbb{E}\{\theta \Upsilon\}$ over the joint distribution of $\theta, \Upsilon$ (with fixed marginal for $\theta$), subject to

$$\mathbb{E}\{\Upsilon^2\} = D_D, \quad \mathbb{E}\{\Upsilon\} = 0, \quad (17)$$

where $\Upsilon$ is defined as $\Upsilon \triangleq X - \mathbb{E}\{X|Y\}$. Next, let us define $\Upsilon_G$ as the Gaussian reconstruction error that satisfies the constraints in (17), i.e., the encoder generates $Y = X + \alpha \theta + T$, where $\alpha$ is a constant and $T \sim \mathcal{N}(0, \sigma_T^2)$ is independent, of $\theta$ and $X$, and the decoder uses the optimal estimator which is linear and hence yields jointly Gaussian reconstruction $\hat{X}$ and reconstruction error $X - \hat{X} \triangleq \Upsilon_G$. Using Lemma 1, we relate the two random variables $\Upsilon, \Upsilon_G$ as

$$\Upsilon = \eta(\Upsilon_G, Z) \quad (18)$$

where $\eta : \mathbb{R} \times \mathbb{R} \to \mathbb{R}$ is a deterministic function and $Z$ is a random variable distributed independently from $\Upsilon_G$, and $Z - (\Upsilon_G, \Upsilon) - \theta$ forms a Markov chain in this order. Hence, the objective can be expressed as: minimize

$$J = \mathbb{E}\{\theta\eta(\Upsilon_G, Z)\} \quad (19)$$

over $\eta(\cdot,\cdot)$ and the joint distribution of $Z$ and $\theta$ which is denoted here as $f_{Z,\theta}(z,\theta)$. Next, we expand $J' = \min_{\eta(\cdot,\cdot), f_{Z,\theta}} J$:

$$J' = \min_{\eta, f_{Z,\theta}} \mathbb{E}\{\theta\eta(\Upsilon_G, Z)\} \quad (20)$$

$$= \min_{\eta, f_{Z,\theta}} \int \mathbb{E}\{\theta\eta(\Upsilon_G, Z)|Z=z\} f_Z(z)\mathrm{d}z \quad (21)$$

$$\geq \int \inf_{\eta, f_{Z,\theta}} \mathbb{E}\{\mathbb{E}\{\theta\eta(\Upsilon_G, Z)|Z=z, \Upsilon_G, \Upsilon\}\} f_Z(z)\mathrm{d}z \quad (22)$$

$$= \min_{f_Z} \int \inf_{\eta_z} \mathbb{E}\{\theta\eta_z(\Upsilon_G)\} f_Z(z)\mathrm{d}z \quad (23)$$

where $\eta_z : \mathbb{R} \to \mathbb{R}$ is a deterministic function that depends on the realization $Z = z$, (22) is due to Fatou's lemma (cf. [35]), and (23) is a consequence of the Markov chain $Z - (\Upsilon, \Upsilon_G) - \theta$, and independence of $Z$ and $\Upsilon_G$. Note that

$$\int \mathbb{E}\{\eta_z^2(\Upsilon_G)\} f_Z(z)\mathrm{d}z = D_D$$
$$\int \mathbb{E}\{\eta_z(\Upsilon_G)\} f_Z(z)\mathrm{d}z = 0 \quad (24)$$

hold due to (17). We can equivalently consider an instance of $Z = z$, where we have $\mathbb{E}\{\eta_z(\Upsilon_G)\} = 0$ and $\mathbb{E}\{\eta_z^2(\Upsilon_G)\}$ is fixed. Noting that $\theta$ and $\Upsilon_G$ are jointly Gaussian, we invoke Lemma 2 to conclude that $\Upsilon = \Upsilon_G$ minimizes (19). Hence, without loss of generality, we take $Y = X + \alpha\theta + T$ where $T \sim \mathbb{N}(0, \sigma_T^2)$ is independent of $X$ and $\theta$. Next, we find the value of $\alpha$ and $\sigma_T^2$ at the equilibrium. We first expand $D_E$ as

$$D_E = \mathbb{E}\{(X + \theta - \hat{X})^2\}$$
$$= \mathbb{E}\{(X+\theta)^2\} - \mathbb{E}\{(2\theta + X)\hat{X}\} - \mathbb{E}\{(X-\hat{X})\hat{X}\} \quad (25)$$

Note that the last term in (25) vanishes due to orthogonality of the MSE error to the reconstruction $\hat{X}$. The first term in (25) is constant with respect to the optimization variables ($\sigma_T^2$ and $\alpha$); hence the objective can be re-expressed as maximizing $\mathbb{E}\{(2\theta + X)\hat{X}\}$, and by replacing the expression for $\hat{X}$ as: maximize

$$J(\alpha, \sigma_T^2) = \left(\frac{1+\alpha\rho}{1+r\alpha^2+2\alpha\rho+\frac{\sigma_T^2}{\sigma_X^2}}\right) \mathbb{E}\{(2\theta+X)(X+\alpha\theta+T)\}$$

$$= \sigma_X^2 \left(\frac{(1+\alpha\rho)(1+2\alpha r+\alpha\rho+2\rho)}{1+r\alpha^2+2\alpha\rho+\frac{\sigma_T^2}{\sigma_X^2}}\right) \quad (26)$$

over $\alpha$ and $\sigma_T^2$. Clearly, the choice of $\sigma_T^2 = 0$ maximizes $J$ irrespective of $\alpha$. The solutions to $\frac{dJ}{d\alpha} = 0$ are

$$\alpha^* = \frac{-1 \pm \sqrt{1+4(r+\rho)}}{2(r+\rho)} \quad (27)$$

Noting that $\frac{d^2 J}{d\alpha^2} > 0$ for $\alpha > 0$ and $\frac{d^2 J}{d\alpha^2} < 0$ for $\alpha < 0$, the global maximizer of $J$ is either on the boundary or the one in (14). Noting $\lim_{\alpha\to-\infty} J(\alpha, 0) < J(\alpha^*, 0)$ we obtain (14). Plugging (14) into (6), and after some algebraic manipulations, we arrive at (15) and (16). $\square$

**Remark 3.** *An interesting aspect of the equilibrium is that $|\alpha| < 1$ for all problem parameters. This implies that although the transmitter wants the receiver to reconstruct $X + \theta$ as its estimate, it does not directly transmit $X + \theta$. A high level interpretation of this observation is that at the equilibrium the transmitter never flat out lies.*

We next focus on the impact of $r$ on $D_D$. We plot the costs as a function of $r$ for $\rho = 0$ in Figure 3a. The following result is a direct consequence of Theorem 4.

**Corollary 1.** *As $r \to \infty$, $D_D \to \sigma_X^2/2$, and as $r \to \rho^2$, $D_D \to 0$.*

The effect of correlation $\rho$ on $D_E$ and $D_D$ is illustrated in Figure 3b. As can be seen in Figure 3b, $D_D$ is an increasing function of $\rho$ while $D_E$ is decreasing in $\rho$, as intuitively expected: if $\rho = -1$, the objective of the transmitter is to make $\hat{X} = 0$ (for $r = 1$), which can be achieved by transmitting nothing. This equilibrium in this case is referred to as the "babbling equilibrium" in the cheap talk literature [22]. As $\rho$ increases, $D_E$ increases as well and the transmitted message becomes more informative for the receiver. At the extremal point of $\rho = 1$, the receiver can reconstruct $X$ perfectly.

### B. Compression

We next characterize $\mathcal{RD}_S$ for general sources and distortion measures.

**Theorem 5.** *$\mathcal{RD}_S$ is the convex hull of the set of all triplets $(R, D_E, D_D)$ for which there exists a (deterministic) function $h : \mathcal{Y} \to \hat{\mathcal{X}}$ and a conditional distribution $p(Y|X,\theta)$ such that*

$$R \geq I(X, \theta; Y)$$

$$D_E \geq \mathbb{E}\{d_E(X, \theta, h(Y))\}, \quad D_D \geq \mathbb{E}\{d_D(X, h(Y))\}.$$

The proof of Theorem 5 directly follows from the standard rate-distortion arguments and is omitted here (see e.g., [16]). Next, we specialize to the quadratic-Gaussian case.

**Lemma 3.** *All equilibrium points of $\mathcal{RD}_S$ are achieved, uniquely, by the jointly Gaussian $(X, Y, \theta)$ triplet.*

The proof follows from the well-known property of Gaussian distribution achieving maximum entropy under a variance constraint and the proof of Theorem 4. The following theorem characterizes the strategic R-D function for the quadratic-Gaussian equilibrium.

**Theorem 6.** *For the Q-G setting of Problem 2, the equilibrium $(D_E, D_D)$ pair in terms of $R$ is:*

$$D_D = \sigma_X^2 2^{-2R}\left(1 + (2^{-2R} - 1)\left(\frac{(r-\rho^2)(A-1)}{A(2r+A\rho+\rho)}\right)\right) \quad (28)$$

$$D_E = \sigma_X^2 \left(1 + 2\rho + r - (1 - 2^{-2R})\frac{A(r+\rho)+\rho}{A-1}\right) \quad (29)$$

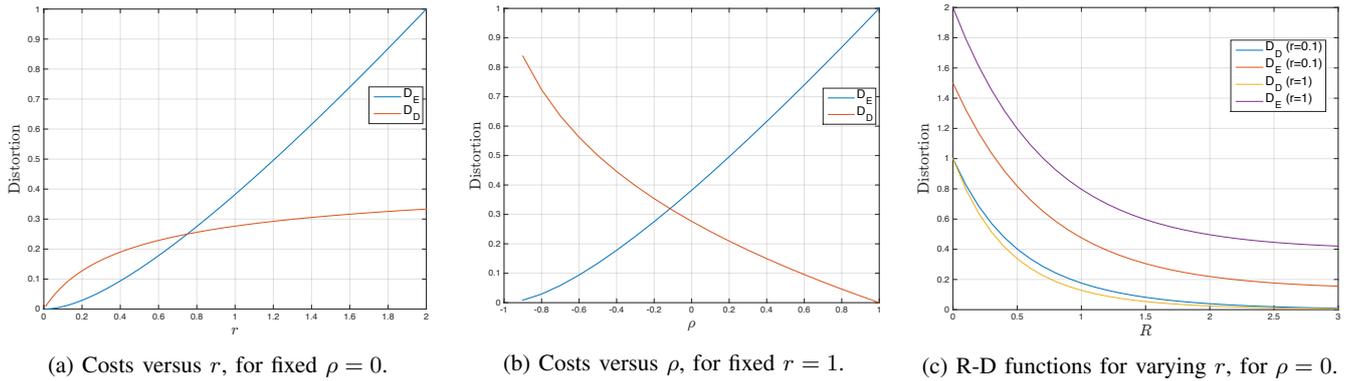

(a) Costs versus $r$, for fixed $\rho = 0$.   (b) Costs versus $\rho$, for fixed $r = 1$.   (c) R-D functions for varying $r$, for $\rho = 0$.

Fig. 3: Numerical analysis of the equilibrium.

where $A = \sqrt{1 + 4(r + \rho)}$.

*Proof.* From Lemma 3, we have $Y = X + \beta\theta + S$ for some $\beta \in \mathbb{R}$ where $S \sim \mathcal{N}(0, \sigma_S^2)$ is independent of $X$ and $\theta$. Plugging this representation in Theorem 5, we obtain the following characterization of $R, D_D, D_E$ in terms of $\sigma_S^2$:

$$R = \frac{1}{2}\log\left(1 + \frac{\sigma_X^2}{\sigma_S^2}(1 + \beta^2 r + 2\beta\rho)\right) \quad (30)$$

$$D_D = \sigma_X^2 \left(\frac{\beta^2(r - \rho^2) + \frac{\sigma_S^2}{\sigma_X^2}}{1 + 2\beta\rho + \beta^2 r + \frac{\sigma_S^2}{\sigma_X^2}}\right) \quad (31)$$

$$D_E = \sigma_X^2\left(1 + 2\rho + r - \frac{(1+\beta\rho)(1+2\beta r + \beta\rho + 2\rho)}{1 + r\beta^2 + 2\beta\rho + \frac{\sigma_S^2}{\sigma_X^2}}\right) \quad (32)$$

Using (44), we have

$$\sigma_S^2 = \sigma_X^2\left(\frac{1 + \beta^2 r + 2\beta\rho}{2^{2R} - 1}\right) \quad (33)$$

We next note that the objective of the encoder is to minimize $D_E$ over the possible choices of $\beta$, which is equivalent to maximizing

$$J(\beta) = \frac{(1+\beta\rho)(1+2\beta r + \beta\rho + 2\rho)}{1 + r\beta^2 + 2\beta\rho}, \quad (34)$$

which is the same expression as (26) with $\beta$ replacing $\alpha$, and with $\sigma_T^2 = 0$ (and $\sigma_X^2 = 1$). The maximizer of $J(\beta)$ then follows from the proof of Theorem 4 as

$$\beta^* = \frac{-1 \pm \sqrt{1 + 4(r + \rho)}}{2(r + \rho)}. \quad (35)$$

Plugging (45) into (31) and (45), and using (35), we arrive at (28) and (29). □

**Remark 4.** *An interesting aspect of strategic Q-G compression is that the forward test channel of the rate-distortion function can be expressed as $Y = X + \beta^*\theta + S$, where $\beta^*$ (given in (35)) is independent of the allowed rate and $S \sim \mathcal{N}(0, \sigma_S^2)$ is independent of $X$ and $\theta$. Moreover $\beta^* = \alpha$, where $\alpha$ is the coefficient in the simple equilibrium, given in Theorem 4. Hence, the problem of strategic compression simplifies to compressing $X + \alpha\theta$. This enables, in practice, the use of standard encoding codes for strategic compression operating on the effective source $X + \alpha\theta$. We note that such simplifications do not carry out to settings with receiver SI, as analyzed in Section IV.*

The strategic R-D functions, for fixed $\rho = 0$, and $r = 1$ or $r = 0.1$, are plotted in Figure 3c.

### C. Equilibrium with Noisy Channel

We next analyze the Q-G equilibrium with noisy channel (described in Section II-E), where we investigate whether the single-letter strategies similar to ones in Theorem 2 continue to have Shannon sense optimality.

**Theorem 7.** *For the equilibrium of Q-G Problem 3 with noisy channel, the strategies*

$$g^*(X, \theta) = \sqrt{\frac{P_T}{\sigma_X^2(1 + 2\alpha\rho + \alpha^2 r)}}(X + \alpha\theta), \quad h^*(Y) = \mathbb{E}\{X|Y\} \quad (36)$$

*with $\alpha = \frac{-1 + \sqrt{1 + 4(r + \rho)}}{2(r + \rho)}$ are Shannon sense optimal for all power levels.*

**Remark 5.** *If a single-letter strategy is Shannon sense optimal, it is also optimal among all single-letter strategies.*

*Proof.* Using data processing inequality [16],

$$R(D_E) \leq C(P_T), \quad (37)$$

one obtains a lower bound on the distortion of any source-channel coding scheme [10, Theorem 21]. The capacity of the AWGN channel is given by

$$C(P_T) = \frac{1}{2}\log\left(1 + \frac{P_T}{\sigma_N^2}\right). \quad (38)$$

The strategic R-D function is given in (29). Plugging (29) and (38) into (37), we obtain

$$D_E \geq \sigma_X^2\left(1 + 2\rho + r - \frac{(1+\alpha\rho)(1+2\alpha r + \alpha\rho + 2\rho)}{(1 + r\alpha^2 + 2\alpha\rho)(1 + \frac{\sigma_N^2}{P_T\sigma_X^2})}\right).$$

Computing distortion associated with the mappings in (36) yields a lower bound on $D_E$:

$$D_E \leq \sigma_X^2 \left(1 + 2\rho + r - \frac{(1+\alpha\rho)(1+2\alpha r + \alpha\rho + 2\rho)}{(1+r\alpha^2+2\alpha\rho)(1+\frac{\sigma_N^2}{P_T\sigma_X^2})}\right).$$

Noting that inner and outer bounds match, we arrive at the desired result. □

**Remark 6.** *Note that the encoding map at the equilibrium does not depend on the channel noise variance. Moreover, paralleling Remark 4, the solution in Theorem 7 also admits an "effective source" interpretation: the optimal encoding strategy is identical to the one in the non-strategic setting stated in Theorem 2, applied to the effective source $X + \alpha\theta$.*

## IV. IMPACT OF RECEIVER SIDE INFORMATION

Next, we extend our analysis to the receiver side information setting as shown in Figure 1. We begin with the noiseless channel case with receiver SI. The focus of our results is the quadratic-Gaussian setting, i.e., $(X, \theta, W) \sim \mathcal{N}(0, R_{X\theta W})$, where $R_{X\theta W}$ is parametrized as

$$R_{X\theta W} = \sigma_X^2 \begin{bmatrix} 1 & \rho_{X,\theta} & \rho_{X,W} \\ \rho_{X,\theta} & r_\theta & \rho_{\theta W} \\ \rho_{X,W} & \rho_{\theta W} & r_W \end{bmatrix}$$

and the distortion measures are given as in (5). The following lemma states that mappings at the equilibrium are linear (affine if variables have non-zero mean).

**Lemma 4.** *The equilibrium for the Q-G noiseless channel setting is achieved by mappings*

$$g(X, \theta) = X + \alpha_{SI}\theta, \quad h(Y, W) = bY + cW, \quad (39)$$

*for some $\alpha_{SI}, b, c \in \mathbb{R}$.*

The proof follows steps identical to those of Theorem 4. The coefficients, $\alpha_{SI}, b, c$ at this equilibrium can be explicitly computed as in the case of Theorem 4, but this computation is rather involved and not included here. Instead, we focus on the high level impact of SI. We first analyze the benefit of the presence of receiver SI at the transmitter side. This question is intimately related to the feedback scenarios in strategic communication: if the receiver has the option of conveying its SI to the transmitter, should it choose to do so? Let us define $D_E^{SI}$ and $D_D^{SI}$ as the distortions of the transmitter and the receiver at the equilibrium with receiver SI; and $D_E^{RSI}$ and $D_D^{RSI}$ as the distortions of the transmitter and the receiver in the setting where SI is also available at the transmitter. The following theorem states that in the Q-G setting, the presence of the receiver SI at the transmitter is not useful to the transmitter or to the receiver.

**Theorem 8.** *In the Q-G setting, the following holds:*

$$D_E^{RSI} = D_E^{SI}, \quad D_D^{RSI} = D_D^{SI}.$$

*Proof.* We begin by showing optimality of linear strategies in the setting where SI is available at both the transmitter and the receiver. First, we constrain the set of encoding strategies so that the encoding strategy takes $(X' \triangleq X - \mathbb{E}\{X|W\}, \theta' \triangleq \theta - \mathbb{E}\{\theta|W\})$ as its arguments (as opposed to $(X, \theta, W)$). Due to the jointly Gaussian statistics, $(X', \theta')$ are statistically independent of $W$. The receiver has also access to $W$ (hence $W$ is common information), and there is no loss of generality imposed by this constraint on the encoding strategy, i.e., the problem is equivalent to the one without any SI with distortion function of the transmitter $\mathbb{E}\{(X' + k\theta' - \hat{X})^2\}$ for some $k \in \mathbb{R}$. The optimality of linear strategies, in $X', \theta'$ then follows from Theorem 4. Since $X', \theta'$ are linear functions of $X, \theta, W$, due to the jointly Gaussian statistics, the encoding strategy is linear in $X, \theta, W$ as well.

Given that the encoding strategy is in the form of $Y = X + a\theta + bW$ (for some $a, b \in \mathbb{R}$), and $W$ is also available at the receiver, $Y = X + a\theta$ and $Y = X + a\theta + bW$ yield identical costs since the receiver can simply subtract $bW$ from $Y$. Hence, there is no need for the transmitter to use $W$ in its encoding strategy. □

**Remark 7.** *We note the absence of the constraint associated with $Y$ in Theorem 8. A constraint on $Y$, such as an average power constraint in the form of $\mathbb{E}\{Y^2\} \leq P$ for some $P \in \mathbb{R}^+$ renders the realization of $W$ useful to the transmitter and also to the receiver as will be shown in Theorem 11.*

We next consider the compression problem in this setting (see [36] for the analogous non-strategic problem). The achievable rate-distortion region $(R, D_E, D_D)$, denoted by $\mathcal{RD}_S^{SI}$, can be characterized following the arguments in [36] and is presented in the following theorem.

**Theorem 9.** $\mathcal{RD}_S^{SI}$ *is the convex hull of the set of all triplets $(R, D_E, D_D)$ for which there exist a function $h: \mathcal{X} \times \mathcal{W} \to \hat{\mathcal{X}}$ and a conditional distribution $p(Y|X, \theta)$ such that*

$$R \geq I(X, \theta; Y) - I(Y; W) \quad (40)$$
$$D_E \geq \mathbb{E}\{d_E(X, \theta, h(Y, W))\} \quad (41)$$
$$D_D \geq \mathbb{E}\{d_D(X, h(Y, W))\} \quad (42)$$

**Remark 8.** *In general IT problems, side information has two types of benefits for the receiver, as demonstrated in Theorem 9 (for a detailed analysis, see [28, Section 11]). The first one is estimation benefit, which corresponds to the receiver using $W$ (in addition to $Y$) to generate $\hat{X}$, as shown in (41) and (42). This benefit also exists in the single-letter case. The second one, namely the rate reduction benefit only exists in the IT setting, and is demonstrated by the term $I(Y; W)$ in (40). In non-strategic settings, the encoder makes $Y$ correlated with $W$ to maximize this rate reduction. However, in strategic settings, there exist problem parameters that render $Y$ independent of $W$ due to the misalignments of the objectives, $d_E$ and $d_D$, and hence make $I(Y; W)$ vanish. This observation plays a pivotal role in the equilibrium with noisy channel and the receiver SI.*

Next, we extend our analysis to the Q-G setting, as shown in Figure 1, where $X, \theta, W$ is jointly Gaussian. The following lemma characterizes the forward test channel that achieves the $\mathcal{RD}_S^{SI}$.

**Lemma 5.** *In the Q-G setting, $\mathcal{RD}_S^{SI}$ is achieved by*

$$Y = X + \beta(R)\theta + S$$

where $S \sim \mathcal{N}(0, \sigma_S^2)$ is statistically independent of $X$, $\theta$ and $W$. The equilibrium coefficient $\beta(R)$ and $\sigma_S^2$ depend on $R$.

*Proof.* The fact that jointly Gaussian $X, \theta, Y$ triple achieves $\mathcal{RD}_S^{SI}$ follows from Lemma 4 and the entropy maximization property of jointly Gaussian distributions subject to second order constraints [37]. From the definition of the problem, we have the natural Markov chain $Y - (X, \theta) - W$ (see e.g., [38]). Hence, we have

$$Y = X + \beta\theta + S \tag{43}$$

for some $\beta \in \mathbb{R}$, and $S \sim \mathcal{N}(0, \sigma_S^2)$ is independent of $X, \theta$ and $W$. Plugging (43) into (40), we have

$$R = \frac{1}{2}\log\left(1 + \frac{\sigma_X^2}{\sigma_S^2}\left(1 + \beta^2 r_\theta + 2\beta\rho_{X\theta} - \frac{(\rho_{XW} + \beta\rho_{\theta W})^2}{r_W}\right)\right), \tag{44}$$

and into (41), we obtain

$$D_E = \sigma_X^2\left(1 + 2\rho_{X\theta} + r_\theta - \frac{(1 + \beta\rho_{X\theta})(\beta^2 r_\theta + 2\beta\rho_{X\theta})}{1 + \beta^2 r_\theta + 2\beta\rho_{X\theta} + \frac{\sigma_S^2}{\sigma_X^2}}\right),$$

noting that $h(Y, W) = \mathbb{E}\{X|Y, W\}$ due to quadratic $d_D$ and is linear due to jointly Gaussian $X, Y, W$. Using (44), we have

$$\frac{\sigma_S^2}{\sigma_X^2} = \frac{1}{2^{2R} - 1}\left(1 + \beta^2 r_\theta + 2\beta\rho_{X\theta} - \frac{(\rho_{XW} + \beta\rho_{\theta W})^2}{r_W}\right) \tag{45}$$

We next note that the objective of the encoder is to minimize $D_E$ over the possible choices of $\beta$, which is equivalent to maximizing

$$J(\beta) = -\frac{(1 + \beta\rho_{X\theta})(\beta^2 r_\theta + 2\beta\rho_{X\theta})}{1 + \beta^2 r_\theta + 2\beta\rho_{X\theta} + \frac{\sigma_S^2}{\sigma_X^2}} \tag{46}$$

over $\beta$. Plugging (45) into (46) we observe that $\beta^* = \arg\max J(\beta)$ depends on $R$. Note that $J$ is continuous in $\beta$ and is bounded above, and hence it admits a maximum by Weierstrass theorem [35]. □

**Remark 9.** *As stated in Remark 4, the compression coefficient (without SI) $\beta^*$ is independent of the allowed rate, and identical to the equilibrium coefficient $\alpha$ in Theorems 4 and 7. Here, due to SI, particularly, the $I(Y; W)$ term, $\beta$ depends on the allowed rate, and is obviously different from $\alpha_{SI}$ in Lemma 4 where there is no rate constraint.*

Next, we analyze the benefit of the presence of SI at the transmitter side. At first sight, it might seem that due to the strategic aspect of the problem at hand, the presence of this SI should help the transmitter. The following theorem states that this intuition is not correct; specifically, there is no benefit of the presence of the receiver SI at the transmitter side.

**Theorem 10.** *In the Q-G setting, the following holds:*

$$\mathcal{RD}_S^{SI} = \mathcal{RD}_S^{RSI}.$$

*Proof.* We first note that, following the arguments in the proof of Theorem 8, the transmitter SI does not affect the distortions ($D_E$ and $D_D$). When SI is available at both ends, it can be shown using the arguments in [39] and Theorem 3 that the rate expression simplifies to $R = \min I(X, \theta; Y|W)$ where minimization is over all conditional probability distributions $p(Y|X, \theta, W)$, while when SI is only available at the encoder we have the same minimization over $p(Y|X, \theta)$. Hence, the only difference is due to the additional Markov chain constraint $Y - (X, \theta) - W$. It is well known that for jointly Gaussian variables this constraint is always satisfied (see e.g., [36]), and hence it does not affect the minimization. □

Finally, we study the Q-G equilibrium with noisy channel and receiver SI. We first investigate the optimal single-letter strategy within the set of linear strategies.

**Lemma 6.** *Optimal linear strategies in the case of the noisy Q-G setting with SI are*

$$g(X, \theta) = \sqrt{\frac{P_T}{\sigma_X^2(1 + 2\alpha_{SI}\rho_{X,\theta} + \alpha_{SI}^2 r_\theta)}}(X + \alpha_{SI}\theta),$$

$$h(Y, W) = \mathbb{E}\{X|Y, W\}$$

*where $\alpha_{SI}$ the same coefficient as in Lemma 4.*

The proof of Lemma 6 follows from standard minimum mean-squared error (MMSE) computations, very similar to the derivation of $D_E$ in the proof of Lemma 5. Next, we present our main result pertaining to this setting.

**Theorem 11.** *In the strategic Q-G setting with channel noise and receiver SI, single-letter linear strategies provided in Lemma 6 are Shannon sense optimal if, and only if,*

$$\rho_{X,W} = -\rho_{\theta,W}\beta(R), \tag{47}$$

*where $R$ is given as*

$$R = \frac{1}{2}\log\left(1 + \frac{P_T}{\sigma_N^2}\right). \tag{48}$$

*Proof.* Equating the outer bound obtained via data processing inequality $R_S^{SI}(D) \leq C(P)$ to the inner bound achieved by the linear mapping in Lemma 6, we get a matching condition which states that for the Shannon sense optimality, the communication channel in Figure 1 must be identical to the R-D test channel provided in Lemma 5[9]. Note that $\alpha_{SI}$ does not depend on the channel parameters $P_T$ or $\sigma_N^2$. However, $\beta(R)$ depends on the rate, and hence on the channel parameters, due to (48). The only way to make the R-D test channel identical to the actual one is to operate at the rate which satisfies $\beta(R) = \alpha$. By Theorem 9, $\beta(R) = \alpha$ implies that $I(Y; W) = 0$, which is equivalent to statistical independence of $Y$ and $W$, which implies uncorrelated variables (due to the joint Gaussian statistics), and hence we have (47). □

**Remark 10.** *Theorem 11 does not preclude the possibility of optimality of the mappings in Lemma 6 within the set of single-letter strategies even if they do not satisfy (47) in which case they are strictly suboptimal in the Shannon sense (i.e., among $n$-letter strategies).*

---

[9]The same matching condition was obtained in [40] for the non-strategic variant of the same problem.

## V. Decentralized Stochastic Control Problems

In this section, we focus on Problem 4. The following theorem extends the main result of [13] to the strategic setting.

**Theorem 12.** *The essentially unique solution to Problem 4 comprises linear mappings $g(X) = c(X + \alpha\theta)$ and $h(Y) = dY$ for some $\alpha, c, d \in \mathbb{R}$, if, and only if, $\varphi_E$ and $\varphi_D$ do not involve any product terms in $U$ and $\hat{X}$.*

**Remark 11.** *Before delving into the proof, we note that the proof does not directly follow from Theorem 7. This is because the cross term in the objective functions, $\mathbb{E}\{U(a_1 X + a_2 \theta)\}$, for some $a_1, a_2 \in \mathbb{R}$, cannot be upper bounded using Cauchy-Schwarz inequality (as done in [13] for the term $\mathbb{E}\{XU\}$), since a linear mapping for $g(X, \theta)$ does not necessarily imply $g(X, \theta) = \kappa(a_1 X + a_2 \theta)$ for some $\kappa \in \mathbb{R}$. Hence, in the proof, we address the problem from the beginning.*

*Proof.* First, we follow very similar steps to ones in [13] to show that the problem is equivalent to the one with

$$\varphi'_E = (X + \theta - \hat{X})^2 + k_1 U^2 + k_2 UX + k_3 U\theta \quad (49)$$
$$\varphi'_D = (X - \hat{X})^2 \quad (50)$$

as the objective functions in the underlying stochastic game. The optimal mapping for the second agent is $h(Y) = \mathbb{E}\{X|Y\}$, and hence, the problem simplifies to optimization over the encoding mapping $g$. Consider the dual (equivalent) problem of minimizing $J_E = \mathbb{E}\{\varphi'_E\}$ subject to a fixed $J_D = \mathbb{E}\{\varphi'_D\}$. Let us expand $J_E$:

$$J_E = D_E + \mathbb{E}\{k_1 U^2 + k_2 UX + k_3 U\theta\}. \quad (51)$$

where

$$D_E = \mathbb{E}\{(X - \mathbb{E}\{X|Y\})^2\} + 2\mathbb{E}\{\theta(X - \mathbb{E}\{X|Y\})\} + \mathbb{E}\{\theta^2\}.$$

We follow the same steps as those in the proof of Theorem 4 to conclude that jointly Gaussian triplet $Y, X, \theta$ minimizes $D_E$ for any given, fixed $J_D$. Let us now consider the original problem of minimizing $J_E$. For any given $J_D$, $D_E$, the problem simplifies to minimization of

$$\mathbb{E}\{k_1 U^2 + k_2 UX + k_3 U\theta\}$$

subject to fixed $D_E$. Here, $\mathbb{E}\{k_1 U^2\}$ corresponds to a power constraint of the form $\mathbb{E}\{U^2\} = P_T$ for some $P_T \in \mathbb{R}^+$, hence the problem can be transformed into a minimization problem with hard constraints: minimize $\mathbb{E}\{U(X + \frac{k_3}{k_2}\theta)\}$ subject to fixed $D_E$ and $\mathbb{E}\{U^2\} = P_T$. The power constraint ensures that the optimal $U$ is zero-mean. Noting that $(X + \frac{k_3}{k_2}\theta)$ is Gaussian, and first and second order moments of $U$ are fixed (power constraint fixes $\mathbb{E}\{U^2\}$), the same steps that led to joint Gaussianity of $X, Y, \theta$ for fixed $J_D$ in the preceding part of the proof, also yield joint Gaussianity of $X, \theta$ and $U$ (and thus $Y$). Hence, for any $J_D$ and $D_E$ constraint, we can find a jointly Gaussian $U, X, \theta$ that minimizes the cross terms which are products of a fixed Gaussian random variable and an optimization variable. The only mapping that yields jointly Gaussian $U, X, \theta$ is linear (or affine if the underlying variables are not zero-mean). The "only if" part follows from the observation that an optimal linear strategy implies the optimality of a linear strategy in the non-strategic setting, by casting $X + k\theta$ as the effective $X$ for some $k \in \mathbb{R}$. This contradicts the "only if" part of Theorem 3. □

## VI. Discussion

In this paper, we have addressed some fundamental strategic communication problems. Tools from information theory have played a key role in the derivation of our results as:

- The Q-G equilibrium entails unique linear optimal strategies. The proof is nontrivial and requires results from probability theory, namely *functional representation lemma* and *maximal correlation measure*, which have been extensively used in information theory.
- The noisy Q-G equilibrium with noise in the channel also admits linear optimal strategies. The proof relies on the fundamental concepts of information theory: data processing inequality, and the notions of rate-distortion and channel capacity [10].
- The Q-G equilibrium with receiver SI admits linear optimal strategies, if there is no channel noise present. For the noisy channel case, it does so for the very specific, matched case of the channel noise, the allowed power and the joint statistics of source-private information-side information. It is well known that the non-strategic analogue of this problem does not admit linear optimal strategies. This sharp contrast between the noiseless and noisy strategic settings highlights the need for detailed information-theoretic analysis for such noisy settings.
- We have identified necessary and sufficient conditions under which general strategic games admit linear optimal strategies, in a similar manner to non-strategic stochastic team decision problems [13].

Some future directions include extensions to multi-dimensional and networked settings, to other (than Gaussian) statistics; and applications of the developed SIT framework to other problem areas (for some preliminary work, see [41] for problems involving privacy constraints).

**Emrah Akyol** (S'03-M'12) received the Ph.D. degree from the University of California at Santa Barbara, in 2011. From 2006 to 2007, he held positions at Hewlett-Packard Laboratories and NTT Docomo Laboratories, both in Palo Alto, CA, where he worked on topics in video compression and streaming. From 2013 to 2014, he was a Postdoctoral Researcher with the Department of Electrical Engineering, University of Southern California. Currently, he is a Postdoctoral Research Associate at the Coordinated Science Laboratory, University of Illinois at Urbana-Champaign. His research interests are on information processing challenges associated with socio-cyber-physical systems. He was the recipient of the 2010 UCSB Dissertation Fellowship, the 2014 USC Provost's Postdoctoral Scholar Training and Travel Award and was an invited participant of the 2015 NSF Early Career Investigators' Workshop on Cyber-Physical Systems and Smart City.

**Cédric Langbort** is an associate professor of Aerospace Engineering (with tenure) at the University of Illinois at Urbana-Champaign (UIUC), where he is also affiliated with the Decision & Control Group at the Coordinated Science Lab (CSL), and the Information Trust Institute. Prior to joining UIUC in 2006, he studied at the Ecole Nationale Superieure de l'Aeronautique et de l'Espace-Supaero in Toulouse (France), the Institut Non-Lineaire in Nice, and Cornell University, from which he received the Ph.D. in Theoretical & Applied Mechanics in January 2005. He also spent a year and a half as a postdoctoral scholar in the Center for the Mathematics of Information at Caltech. He works on applications of control, game, and optimization theory to a variety of fields; most recently to "smart infrastructures" problems within the Center for People & Infrastructures which he co-founded and co-directs at CSL. He is a recipient of the NSF CAREER Award, the advisor of a IEEE CDC best student paper award recipient, has been a subject editor for OCAM, the journal of Optimal Control Applications and Methods, and is currently an associate editor for Systems & Control Letters.

**Tamer Başar** (S'71-M'73-SM'79-F'83-LF'13) is with the University of Illinois at Urbana-Champaign (UIUC), where he holds the academic positions of Swanlund Endowed Chair; Center for Advanced Study Professor of Electrical and Computer Engineering; Research Professor at the Coordinated Science Laboratory; and Research Professor at the Information Trust Institute. He is also the Director of the Center for Advanced Study. He received B.S.E.E. from Robert College, Istanbul, and M.S., M.Phil, and Ph.D. from Yale University. He is a member of the US National Academy of Engineering, and Fellow of IEEE, IFAC and SIAM, and has served as president of IEEE CSS, ISDG, and AACC. He has received several awards and recognitions over the years, including the highest awards of IEEE CSS, IFAC, AACC, and ISDG, the IEEE Control Systems (Field) Award, and a number of international honorary doctorates and professorships. He has over 750 publications in systems, control, communications, and dynamic games, including books on non-cooperative dynamic game theory, robust control, network security, wireless and communication networks, and stochastic networked control. He was the Editor-in-Chief of Automatica between 2004 and 2014, and is currently editor of several book series. His current research interests include stochastic teams, games, and networks; security; and cyber-physical systems.